\documentclass[11pt,a4paper,reqno]{amsart}
\usepackage[utf8]{inputenc}

\usepackage{fullpage}
\usepackage{amsaddr}
\usepackage{amssymb}
\usepackage{amsmath}
\usepackage{tikz}
\usepackage{upgreek}
\usepackage{cite}
\newcommand\Tstrut{\rule{0pt}{2.6ex}}         
\newcommand\Bstrut{\rule[-0.9ex]{0pt}{0pt}}

\newcommand{\R}{\mathbb{R}}
\newcommand{\p}{\mathbb{P}}
\newcommand{\E}{\mathbb{E}}

\theoremstyle{plain}

\theoremstyle{definition}

\theoremstyle{remark}

\title{}
\author{}
\date{\today}

\begin{document}
\title[]{Stochastic 3D  modeling of nanostructured NVP/C active material particles for sodium-ion batteries}

\author[]{Matthias Neumann$^{1, \star}$, Tom Philipp$^2$, Marcel Häringer$^3$, Gregor~Neusser$^2$, Joachim R. Binder$^3$, Christine~Kranz$^2$}

\address{$^1$Institute of Stochastics, Ulm University, Helmholtzstraße 18, 89069~Ulm, Germany \\
$^2$ Institute of Analytical and Bioanalytical Chemistry, Ulm University, Albert-Einstein-Allee~11, 89081~Ulm, Germany\\
$^3$Institute for Applied Materials, Karlsruhe Institute of Technology, Hermann-von-Helmholtz-Platz 1, 76344 Eggenstein-Leopoldshafen, Germany 
}

\address{$^\star$Corresponding author, email: matthias.neumann@uni-ulm.de}

\keywords{Electron microscopy; nanostructure; sodium-ion batteries; stochastic structure modeling; structure-property-relationships}
\subjclass{62H11; 82D30; 94A04}

\begin{abstract}
A data-driven modeling approach is presented to quantify the influence of morphology on effective properties in nanostructured sodium vanadium phosphate $\mathrm{Na}_3\mathrm{V}_2(\mathrm{PO}_4)_3$/ carbon composites (NVP/C), which are used as cathode material in sodium-ion batteries. This approach is based on the combination of advanced imaging techniques, experimental nanostructure characterization and stochastic modeling of the 3D nanostructure consisting of NVP, carbon and pores. By 3D imaging and subsequent post-processing involving image segmentation, the spatial distribution of NVP is resolved in 3D, and the spatial distribution of carbon and pores is resolved in 2D. Based on this information, a parametric stochastic model, specifically a Pluri-Gaussian model, is calibrated to the 3D morphology of the nanostructured NVP/C particles. Model validation is performed by comparing the nanostructure of simulated NVP/C composites with image data in terms of morphological descriptors which have not been used for model calibration. Finally, the stochastic model is used for predictive simulation to quantify the effect of varying the amount of carbon while keeping the amount of NVP constant. The presented methodology opens new possibilities for a ressource-efficient optimization of the morphology of NVP/C particles by modeling and simulation. 
\end{abstract}

\maketitle

\section{Introduction}

For energy storage in consumer electronics, electric vehicles and stationary applications, lithium-ion batteries allowing for high energy densities are the dominant technology nowadays~\cite{xiao.2023}. Due to the potential increase in the price of lithium, the toxicity of cobalt required for the widely used active material $\mathrm{LiNi}_{1-x-y}\mathrm{Mn}_x\mathrm{Co}_y\mathrm{O}_2$ (NMC), and the limited supply of both, lithium and cobalt~\cite{jaffe.2017}, alternative battery chemistries are explored. Among these alternatives, sodium-ion batteries~\cite{hwang.2017} are promising candidates. This requires the manufacturing of appropriate electrode materials for an optimized performance of these batteries.  
 
A popular active material for cathodes in sodium-ion batteries is $\mathrm{Na}_3\mathrm{V}_2(\mathrm{PO}_4)_3$ (NVP)~\cite{zhang.2019}. The main benefit of this material arises from its high energy density, cycle stability, and
rate capability~\cite{zheng.2018}. To obtain a sufficient conductivity of the cathodes NVP/carbon composite coatings are applied~\cite{akcay.2021, ling.2018}. The performance of the cell strongly depends on effective properties of the composite material at the nano-scale, which--in turn--are influenced by the spatial distribution of the constituents of this composite material. Optimizing these electrode materials requires a comprehensive understanding of relationships between parameters of the synthesis process, the morphology of the composite material, and the corresponding effective properties.

As shown in \cite{cadiou.2020, furat2021a, kroll.2021, osenberg.2022, wagner.2020} for different electrode materials used in lithium ion batteries, 3D imaging with a subsequent analysis based on methods from spatial statistics and mathematical morphology~\cite{chiu.2013, jeulin.2021} is a powerful tool. It allows for quantitatively studying the nanostructure and their dependence on the synthesis process on the one hand and their impact on effective properties on the other hand. However, merely investigating experimental image data is limited due to the effort of sample preparation, image acquisition and analysis. This problem can be overcome by data-driven parametric stochastic 3D structure modeling, where virtual structures are generated that are statistically similar to the observed ones. By varying the parameters of the stochastic model, virtual structures can be simulated which exhibit different morphological properties than the structures reconstructed based on experimental imaging techniques. To mention two examples, the impact of calendering on the morphology of graphite anode electrodes~\cite{prifling.2019} and structure-property-relationships of solid oxide fuel cell anodes (SOFC) anodes~\cite{moussaoui.2018} have been quantitatively explored with this methodology. Moreover, a stochastic structure model has been developed to generate virtual, but realistic nanostructured NMC particles~\cite{neumann.2023} and has been used to study process-structure-property relationships of the latter.

In the present paper, a stochastic 3D structure model for nanoporous NVP/C active material particles is calibrated based on image data aqcuired by focused ion beam (FIB) - scanning electron microscopy (SEM) tomography, FIB cross-sectioning, transmission electron microscopy (TEM) and energy-dispersive X-ray spectroscopy (EDX). A major challenge in electron microscopy of composite electrodes is the porous nature of the specimen which introduces imaging artefacts, e.g. shine-through artefacts and brightness variations due to the edge effect. This leads afterwards to a demanding analysis and segmentation of the obtained images. To cope with those difficulties an elaborate sample preparation, like embedding in epoxy or silicone resins to fill the pore space and thereby minimize imaging artefacts can be applied prior imaging~\cite{ender.2011} or for instance algorithm-based segmentation approaches by optical flow estimations are deployed~\cite{moroni.2020}. Therefore, an embbeding approach with a silicone resin is used to fill the pore space and enhance contrast between the three phases, NVP, carbon and pore space. Further, to fine tune the segmentation of the detailed carbon and pore space additional high resolution SEM (HR-SEM) images are used. For underpinning of the information obtained by FIB-SEM and HR-SEM, further EDX and TEM measurements were performed. The three-phase nanostructure is modeled by the Pluri-Gaussian model introduced in~\cite{n.2019d}, where we present a methodological framework to consistently combine the information from FIB-SEM and HR-SEM data. The model presented in~\cite{n.2019d} has the advantage that the parameters of the 3D model can be predicted based on 2D SEM data. Note that the required assumption of spatial isotropy is fulfilled in our case. The model is validated by means of morphological descriptors that have not been used for model calibration. Finally, the stochastic model is used for predictive simulation in order to quantify the effect of varying the amount of carbon while keeping the amount of NVP constant.

\section{Materials and imaging}\label{sec:materials_imaging}

\subsection{Synthesis of nanostructured NVP/C particles}\label{sec:synthesis}

For the synthesis of the NVP/C-composite material, $\mathrm{NH}_4\mathrm{VO}_3$, $\mathrm{NH}_4\mathrm{H}_2\mathrm{PO}_4, \mathrm{Na}_2\mathrm{CO}_3$ in molar ration of 4:6:3 were dissolved in water at $73~^\circ$C. Lactose was added as carbon source with 18 mass \%. The resulting solution was spray-dried with an inlet-temperature of $210~^\circ$C and an outlet temperature of $112~^\circ$C. The so obtained precursor were calcined under an argon atmosphere. Hereby, the material was heated up to $450~^\circ$C with a heating rate of 1 K/min. Then the material was heated up to $850~^\circ$C with a heating rate of 3 K/min and a subsequent dwell time of 5 h. Finally, the cooling was done with a cooling rate of 5 K/min to room temperature. This pre-calcined powder was ground in an agitator ball mill for 450 min at 3000 rpm with the use of $\mathrm{ZrO}_2$-milling balls ($\varnothing~0.2$~mm) in an aqueous suspension. Polyacrylic acid (PAA) ($\mathrm{M}_\mathrm{n}$=1800) and polyethylenglycol (PEG) was added to the suspension and it was spray-dried with same conditions. Finally, a second calcination step at $800~^\circ$C (dwell time=5 h, heating rate=3 K/min, atmosphere=$\mathrm{Ar}/\mathrm{H}_2$ (97:3)) is done. Volume fractions of $55.6~\%, 12.6~\%, 31.8~\%$ are calculated for NVP, carbon and pores, respectively, based on the following values: specific pore volume of 0.162~cm$^3$/g, carbon content of 10.5~wt$\%$, a density of 2.88 g/cm$^3$ for the synthesized NVP/C composite, and a density of 3.16~g/cm$^3$ for the pure NVP phase.

\subsection{Sample preparation for electron imaging}

To enhance the contrast between the electrode components and the pore space during the imaging process, the pore space was infiltrated with a silicone resin (ELASTOSIL\textsuperscript{\textregistered} RT601 A/B, Wacker Chemie AG, Germany). A small portion of the NVP/C composite electrode material was immersed into the resin and two vacuum steps were applied (5 min at 240 mbar, 5 min at ambient pressure, followed by 30 min at 240 mbar). Afterwards, the sample was removed from the resin and cured for 24 hours at ambient conditions. Subsequently, the sample was embedded in epoxy resin (EpoFix, Struers GmbH, Germany). The embedded electrode sample was exposed using SiC abrasive papers and the surface polished using diamond suspensions (monocrystalline, 3 µm and 1 µm, Leco Corporation, USA) on Nylon paper (Leco Corporation, USA). Finally, to avoid charging effects during electron imaging the samples were sputtered with an approx. 10 nm thick platinum layer.

\subsection{FIB/SEM, TEM and EDX analysis}\label{sec:imaging}
The FIB/SEM analysis was done with a Helios Nanolab 600 (Thermo Fisher Scientific Inc., USA) with a gallium ion beam operated at 30 kV and beam currents chosen accordingly. Protective platinum layers were deposited at the region of interest via ion beam induced deposition (IBID) using methylcyclopentadienyl trimethyl platinum ($\mathrm{C}_9\mathrm{H}_{16}\mathrm{Pt}$) as precursor prior cross-sectioning and FIB/SEM tomography. 
FIB/SEM tomography of a single particle of the active material was obtained using the “slice and view” software package (Thermo Fisher Scientific Inc., USA). Electron image acquisition was done at 3 kV and 86 pA using immersion mode and the through the lens detector (TLD) operated in backscattering electron (BSE) mode, electron images have a pixel size of $16.6 \times 16.6~\mathrm{nm}^2$. For each image a slice of 20 nm thickness was removed at the front, 200 of these slices were done.
For further phase identification and fine tuning of the segmentation algorithm, FIB/SEM cross-sectioning and HR-SEM imaging was done on another particle of the same active material. TLD BSE images were taken with 3 kV, 0.34 nA and a pixel size of $5 \times 5~\mathrm{nm}^2$. In addition, a lamella with an average thickness of 130 nm was prepared from the same particle and transferred to a copper TEM grid using an OmniProbe manipulation needle (Oxford Instruments plc, GB) and platinum deposition via IBID. 
The TEM lamella was imaged using a TEM JEM-1400 (JEOL Ltd., Japan) operated at 120 kV. EDX mapping was done at the same lamella using a Quanta 3D FEG (Thermo Fisher Scientific Inc., USA) operated at 20 kV and 23 nA equipped with a SDD Apollo XV detector using the Genesis 6.5 acquisition software package (both EDAX Inc., Germany). For the element mapping a size of $512 \times 400$ pixel was chosen, pixel size was $8 \times 8 ~\mathrm{nm}^2$ and the dwell time was 200~$\upmu \mathrm{s}$ with 64 measured frames. The results of EDX measurements are provided as supplementary information, see Figure~\ref{fig:EDX}. In accordance with the EDX analysis the grayscale values in the SEM images can be allovcated as follows: high/bright indicates NVP, medium/gray indicates filled pore space, low/dark indicates the carbon domain.


\begin{figure}[h]
    \centering
    \begin{tabular}{ccc}
         \includegraphics[width = 0.3 \textwidth]{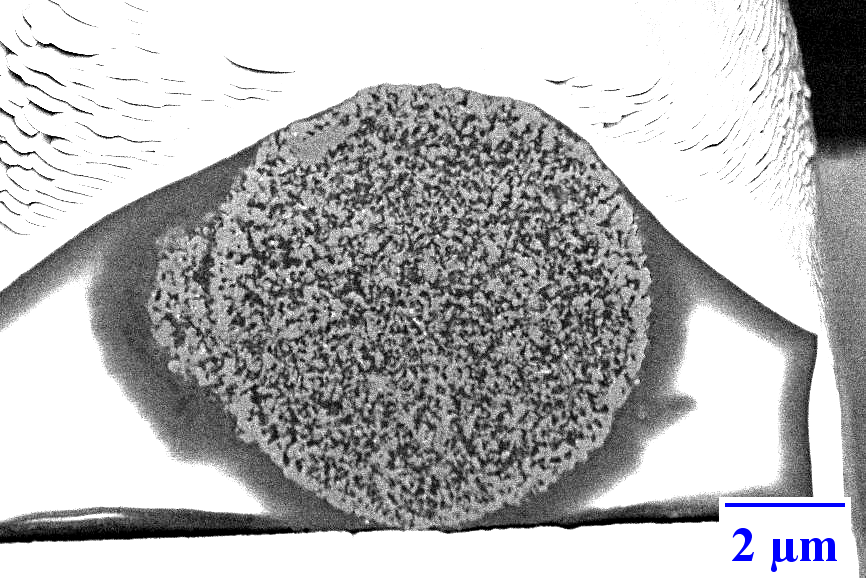} &
         \includegraphics[width = 0.3 \textwidth]{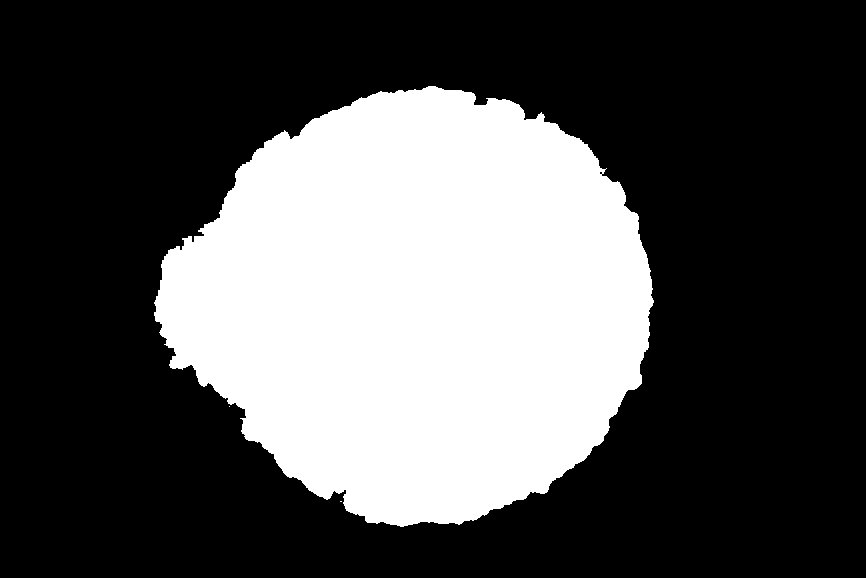}
         &  \includegraphics[width = 0.3 \textwidth]{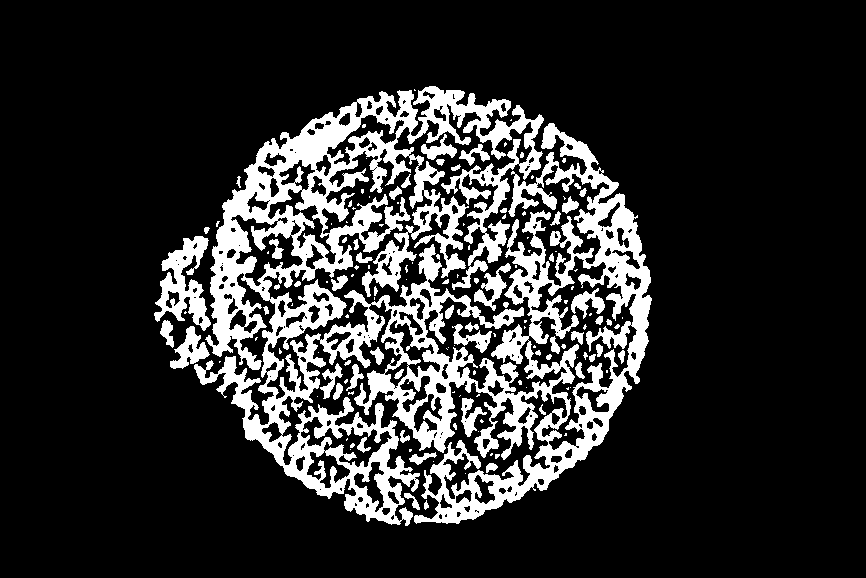} \\ (a) & (b) & (c)
    \end{tabular}
    \caption{BSE image of a cross-section of a NVP/C composite particle (a), area of the NVP/C particle extracted from the BSE image in white using the software Ilastik (b) and the segmented NVP phase (white, c)}
    \label{fig:segmentation_fib}

\end{figure}
\subsection{Image segmentation}\label{sec:image_segmentation}


First, to segment the NVP phase in the image stack obtained by FIB/SEM tomography (exemplary cross-section is shown in Figure 2a), the area of the NVP/C composite particle in the individual images is separated from its surroundings using the software package Ilastik~\cite{sommer.2011}. In short, a random forest classifier is trained using a few hand-labelled voxel and thereby, the area of the particle is separated from the rest of the image (Figure 2b). Within this separated area a local Otsu threshold~\cite{otsu.1979} is applied with a radius of $0.5~\upmu \mathrm{m}$. This procedure is performed on each individual cross-section to compensate for brightness variations within the image stack. Afterwards, a majority filter using a 3 x 3 x 3 voxel neighborhood was applied on the segmented image stack which leads to a smoothing between subsequent slices of the stack.


\begin{figure}[h]
    \centering
    \begin{tabular}{ccc}
         
         \includegraphics[width = 0.27\textwidth]{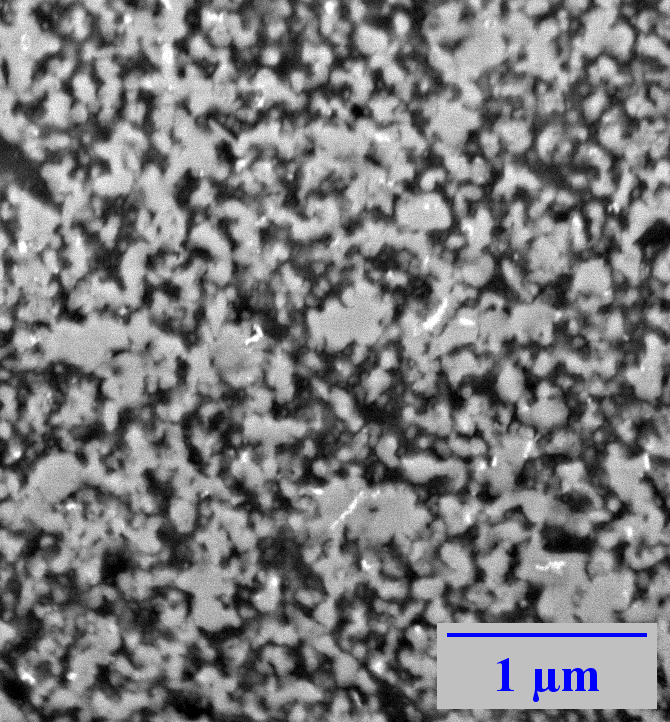}
         & \includegraphics[width = 0.27\textwidth]{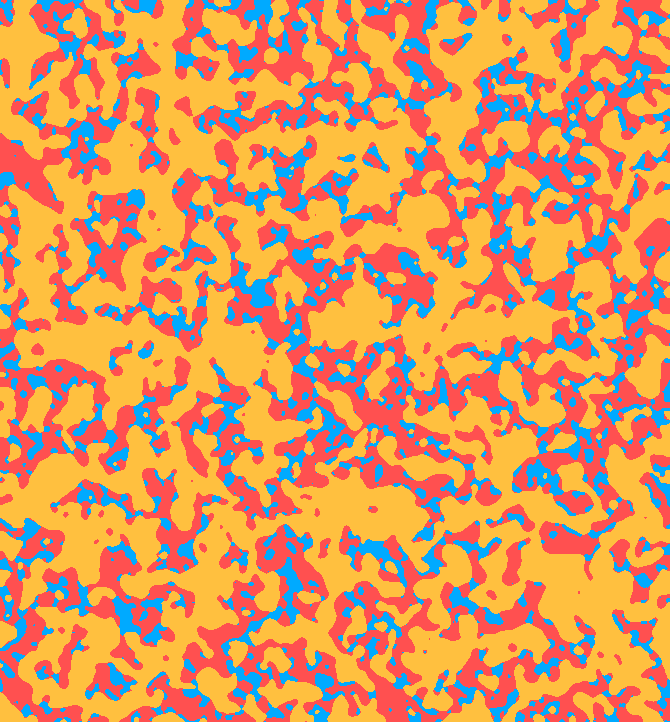} &  
         \includegraphics[width = 0.27 \textwidth]{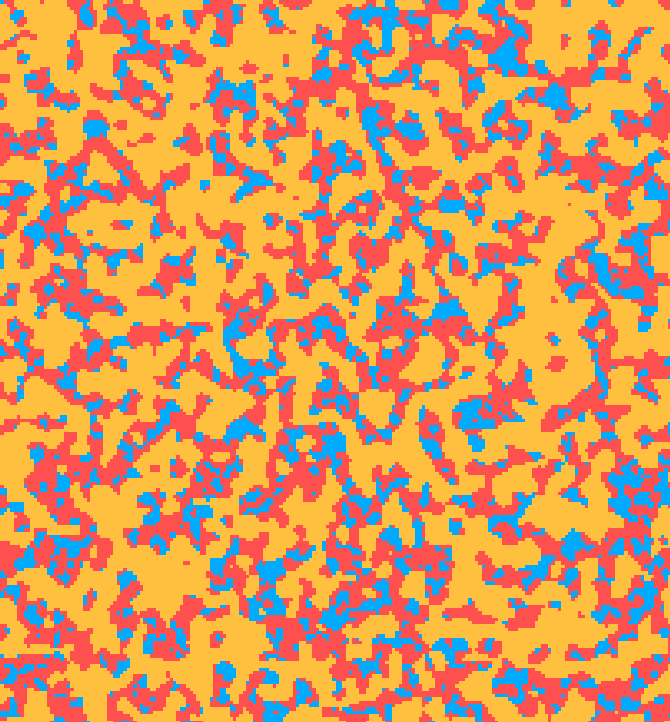}
         \\
        (a) & (b) & (c)
    \end{tabular}
    \caption{Comparison between greyscale image obtained by 2D SEM (a), the corresponding segmented image data (b) and a virtual nanostructure generated by the stochastic 3D model (c). NVP, carbon, and pores are represented in yellow, red and blue, respectively.}\label{fig:model_data_comparison}
\end{figure}

Since the phase identfication between carbon and filled pore space is not unambiguous in the image stack obtained by FIB/SEM tomography, additional HR-SEM images of cross-sections were taken into account. Firstly, a median filter using a radius of 25 nm is applied for noise reduction and a local Otsu threshold~\cite{otsu.1979} is used with a radius of $0.25~\upmu \mathrm{m}$ for segmentation of the NVP phase. In a second step, the complementary area of the image is further subdivided into carbon and filled pore space. Therefore, a $7 \times 7$ Mexican hat filter~\cite{marr.1980} is applied to enhance the contrast at the boundaries of the NVP phase to reduce a missclassification of medium high grayscale values, a usual problem in multi-phase image segmentation~\cite{schluter.2014}. Finally, an Otsu threshold~\cite{otsu.1979} is applied with a radius of $0.25~\upmu \mathrm{m}$ in this complement followed by a majority filter with a radius of 25 nm. This segmentation procedure leads to a good agreement between HR-SEM images and segmentation of the three phases (Figure 3a and 3b).

 \section{Stochastic 3D modeling}\label{sec:stochastic_modeling}

The nanostructure consisting of the active material NVP, carbon and pores is modeled by a Pluri-Gaussian model, i.e., by excursion sets of Gaussian random fields~\cite{adler.2009, chiu.2013}. Excursion sets of Gaussian random fields are an appropriate model for the morphology of electrode materials~\cite{abdallah.2016, moussaoui.2018, n.2019d, prifling.2021b}. In the present paper, we make use of the stochastic model described in \cite{n.2019d}. The model is briefly described in Section~\ref{sec:model_description}. Model calibration to the image data, segmented as presented in Section~\ref{sec:image_segmentation}, is performed in Section~\ref{sec:model_fitting}.

\subsection{Model description}\label{sec:model_description}
 
We consider two motion invariant, i.e., stationary and isotropic, Gaussian random fields $X=\{X(u)\colon u\in \R^3\}$ and $Y=\{Y(u)\colon u\in \R^3\}$ which are centered and standardized. The latter means that the conditions $\mathbb{E}X(u)=0$ and $\mathsf{Var} X(u)=1$ for the expectation and variance of the random variable $X(u)$ are fulfilled for each $u\in\R^3$. Let $\rho_X, \rho_Y \colon \R^3\times \R^3 \rightarrow \R$ denote the covariance functions of $X$ and $Y$, i.e., $\rho_X(u,v) = \mathsf{Cov}(X(u), X(v))$ for all $u,v\in \R^3$ and analogous for $Y$. Note that by the motion invariance of $X$ and $Y$, the values of their covariance functions at a pair $(u,v) \in \R^3\times \R^3$  only depend on the distance $\lvert u-v\rvert$ for all $u,v\in \R^3$. Hence, with some abuse of notation, we write $\rho_X(h)=\rho_X(u,v)$ and $\rho_Y(h)=\rho_Y(u,v)$ for any $h\in[0,\infty)$, where $u,v\in\R^3$ are arbitrary with $h=\lvert u-v\rvert$. By means of the random fields $X$ and $Y$, we define a third random field $Z=\{Z(u)\colon u\in \R^3\}$ by $Z(u) = \sqrt{m} X(u) \, + \sqrt{1-m} \, Y(u)$ for each $u \in \R^3$ and some model parameter $m \in [0,1].$ By construction, $Z$ is also a motion-invariant, centered and standardized Gaussian random field. It inherits these properties from $X$ and $Y$. Note that if $m=0$, the random field $Z$ coincides with $Y$ and, thus, $X$ and $Z$ are independent. The closer the value of $m$ is to 1, the stronger is the correlation between $X$ and $Z$ and in the extreme case $m=1$, the random fields $X$ and $Z$ coincide.

Based on the random fields $X$ and $Z$, we construct random sets $\Xi_1, \Xi_2$ and $\Xi_3$, which denote NVP, carbon, and pores, respectively. First, NVP is modeled as an excursion set of the random field $X$, namely $\Xi_1 = \{u \in \R^3 \colon X(u) \geq \lambda_X \}$ for some parameter $\lambda_X \in \R.$ In a second step, the remaining space is subdivided in carbon and pores with the random field $Z$. We define $\Xi_2 = \lbrace u \in \R^3 \colon X(u) < \lambda_X, Z(u) \geq \lambda_Z \rbrace$ for some model parameter $\lambda_Z \in \R.$ The pore space $\Xi_3 = \R^3 \setminus (\Xi_1 \cup \Xi_2)$ is then given by complement of NVP and carbon. The model is uniquely determined by the two thresholds $\lambda_X, \lambda_Z$, the covariance functions $\rho_X, \rho_Y$ of the underlying Gaussian random fields $X, Y$, and the parameter $m$. By means of the two thresholds, the volume fractions of the two phases can be controlled, while the covariance functions strongly influence the shape of the phases. In particular, the faster the covariance functions decay, the finer is the morphology of the nanostructure in the model. The parameter $m$ controls the dependence between the random fields $X$ and $Z$ and thus it influences the correlation between the phases $\Xi_1$ and $\Xi_2$. For larger values of $m$, the accumulation of $\Xi_2$ around $\Xi_1$ is stronger, i.e., more carbon is accumulated close to the active material.

 \subsection{Model calibration}\label{sec:model_fitting}

 The model calibration is based on analytical relationships between morphological descriptors which can be estimated from image data on the one hand and the levels $\lambda_X, \lambda_Z$, the correlation functions $\rho_X, \rho_Y$ and the parameter $m$ on the other hand. In the present paper, model calibration differs from the approach in~\cite{n.2019d} in the sense, that we use two different types of image data with different resolutions. More precisely, we use 3D image data obtained by FIB-SEM tomography with a voxel size of $16~\mathrm{nm} \times 16~\mathrm{nm} \times 20~\mathrm{nm}$ to fit the covariance function $\rho_X$ which determines the shape of the NVP phase. For calibrating the other parameters, which also require information about the spatial distribution of carbon and the pore space, we use the HR-SEM images with a pixel size of $5~\mathrm{nm} \times 5~\mathrm{nm}$.  

 From image data, we estimate the volume fractions and the two-point coverage probability functions, see Section~3.1.6 in \cite{chiu.2013}. The volume fraction $0 \leq \varepsilon_i \leq 1$ of $\Xi_i$ is defined by 
 \begin{equation}
     \varepsilon_i = \frac{1}{\nu_3(W)}\E [ \nu_3(\Xi_i \cap W)],
 \end{equation}
 for each $1\leq i \leq 3$, where $\nu_3$ denotes the three-dimensional Lebesgue measure and $W \subset \R^3$ with $0 < \nu_3(W) < \infty$ is an arbitrary sampling window. Note that in the case of stationarity, which is assumed for the nanostructures considered in the present paper, the definition of the volume fraction does not depend on the choice of $W$. To estimate the volume fractions from image data, we use the information from 3D FIB-SEM data to determine an estimator $\widehat \varepsilon_1 = 0.5148$ for the volume fraction $\varepsilon_1$ of NVP by voxel counting. As described in Section~\ref{sec:image_segmentation}, carbon can not be distinguished from the pore space in FIB-SEM data. Thus, to estimate the volume fraction of carbon and pores, we make use of the segmented HR-SEM images. Note that by voxel counting, volume fractions of a three-phase material can be estimated from cross-sections in an unbiased way, see Section~10.2.2 in~\cite{chiu.2013}. Doing so, we obtain the estimators $\widehat \varepsilon_{2,\mathrm{2D}} = 0.3328$ and $\widehat \varepsilon_{3,\mathrm{2D}} = 0.1461$ for $\varepsilon_2$ and $\varepsilon_3,$ respectively. For consistency, it is required that $\widehat \varepsilon_1 + \widehat \varepsilon_2 + \widehat \varepsilon_3 = 1$. Thus, based on the 3D estimator for the volume fraction of NVP and the 2D estimators for the volume fraction of carbon and porosity, we define the final estimators $\widehat \varepsilon_2$ and $\widehat \varepsilon_3$ for $\varepsilon_2$ and $\varepsilon_3$ by 
 \begin{equation}
     \widehat \varepsilon_2 = (1 - \widehat \varepsilon_1) \, \frac{\widehat \varepsilon_{2,\mathrm{2D}}}{\widehat \varepsilon_{2,\mathrm{2D}} + \widehat \varepsilon_{3,\mathrm{2D}}}  
 \end{equation} and 
  \begin{equation}
     \widehat \varepsilon_3 = (1 - \widehat \varepsilon_1) \, \frac{\widehat \varepsilon_{3,\mathrm{2D}}}{\widehat \varepsilon_{2,\mathrm{2D}} + \widehat \varepsilon_{3,\mathrm{2D}}} .
 \end{equation}
 This means that we complement the information from 3D image data with the information regarding the ratio of volume fraction of carbon and of porosity. Thereby, we obtain $\widehat \varepsilon_1 = 0.5148, \widehat \varepsilon_2 = 0.3372 $, and $\widehat\varepsilon_3 = 0.1480$. The estimated volume fraction of NVP is in good accordance with the determined value in Section~\ref{sec:synthesis}. The slight underestimation can be attributed to the fact that we consider a cutout from the inner structure of an NVP/C particle and the volume fraction of NVP is larger at the boundary, see Figure~\ref{fig:segmentation_fib}. The mismatch between the estimated volume fractions of carbon and pores compared to the values in Section~\ref{sec:synthesis} arises due to possible small nanopores which are not resolved in the segmentation of HR-SEM images as observed in TEM data, the visualization of which is provided as supplementary information, see Figure~\ref{fig:tem}. Thus, when speaking about the carbon phase, we mean the carbon phase including small nanopores. When discussing the influence of the volume fraction of carbon on the morphology of NVP/C particles in Section~\ref{sec:predictive_simulation}, we provide a formula to recompute the mere volume fraction of carbon based on the volume fraction of carbon including the small nanopores. The latter is reflected in $\widehat \varepsilon_2.$

 In addition to the volume fractions of phase, we use the two-point coverage probability functions $C_{ij}: [0,\infty) \rightarrow [0,1]$ defined by 
 \begin{equation}
     C_{ij}(h) = \p(s \in \Xi_i, t \in \Xi_j),
 \end{equation}
 for $1 \leq i \leq j \leq 3$ and each $h \geq 0$ where $s,t \in \R^3$ such that $|s- t| = h$. Note that due to the assumption of stationarity and isotropy, the values of $C_{ij}(h)$ do not depend on the particular choice of $s$ and $t$. The two-point coverage probabilities indicate spatial dependencies within the considered structure. If the phases at locations having a distance $h$ between each other are independent, then 
 \begin{equation}
      C_{ij}(h) = \p(s \in \Xi_i) \p(t \in \Xi_j) = \varepsilon_i \varepsilon_j,
 \end{equation} where the latter equality follows from Equation (6.34) in~\cite{chiu.2013}. Provided that $C_{ij}(h) \geq \varepsilon_i\varepsilon_j$, then it is more likely that a predefined point belongs to the $i$th phase if there is another point with distance $h$ that belongs to the $j$th phase. An analogous statement can be made if  $C_{ij}(h) < 
 \varepsilon_i \varepsilon_j.$ For computing estimators $\widehat C_{ij}$ for the two-point coverage probability functions $C_{ij}$ for all $1 \leq i \leq 3$ from image data, we use the algorithm described in Section~6.2 of~\cite{ohser.2009} which is based on the fast Fourier transformation. We estimate $C_{11}$, i.e., the two-point coverage probability function of NVP based on all FIB-SEM cross-sections, while the other functions $C_{ij}$ with $j \geq 2$ are estimated based on the HR-SEM images, for which a three-phase segmentation is possible.

 Based on this information contained in the volume fractions and the two-point coverage probability functions, the model parameters are estimated. First, the model parameter $\lambda_X$ is estimated using the relationship $\varepsilon_1 = P(X(u) \geq \lambda_X),$ see Equation (6.157) in~\cite{chiu.2013}, i.e., the estimator $\widehat \lambda_X$ for $\lambda_X$ is given by 
 \begin{equation}\label{eq:epsilon1}
     \widehat \lambda_X = \Phi^{-1}(1 - \widehat \varepsilon_1),
 \end{equation}
 where $\Phi$ denotes the probability distribution function of the standard normal distribution. Furthermore, for all $s,t \in \R$,  let $\varphi_2(s,t,\gamma)$ denote the bivariate probability density function of a two-dimensional Gaussian random vector evaluated at $(s,t)$, where the marginal distributions are standard normal and the correlation coefficient is $-1 \leq \gamma \leq 1.$ With the estimator $\widehat C_{11}$ for $C_{11}$, we determine a non-parametric estimator $\widehat \rho_X$ for $\rho_X$ by means of the relationship 
 \begin{equation}\label{eq:C11}
 C_{11}(h) = \varepsilon_1^2 + \frac{1}{2 \pi}\int_0^{\rho_X(h)} \varphi_2(\lambda_X, \lambda_X, \gamma) \, \mathrm{d}\gamma,
 \end{equation}
 which is valid for each $h>0$. Considering $\widehat \rho_{X}$, it turns out that \begin{equation}\label{eq:choice_rhoX}
     \rho_{X}(h) = \exp(- (\alpha_X h)^2)
 \end{equation} is an appropriate model for the correlation function of the random field $X$. The estimator $\widehat \alpha_X$ for the parameter $\alpha_X$ is determined via the method of least squares. 


 For estimating the remaining parameters, we make use of the following analytical relationships derived in~\cite{n.2019d}, where $\varphi(t)$ denotes the density of the standard normal distribution evaluated at $t \in \R$. The first, which reads as  
 \begin{equation}\label{eq:epsilon2}
     \varepsilon_2 = \int_{-\infty}^{\lambda_X} \varphi(t) \left( 1 - \Phi \left( \frac{\lambda_Z - \sqrt{m}t}{\sqrt{1-m}}\right)\right) \, \mathrm{d}t,
 \end{equation}
 related the volume fraction of the carbon phase to the model parameters $\lambda_Z$ and $m$. The second and the third, given by 
 \begin{equation}\label{eq:C22}
     C_{22}(h) = \int_{-\infty}^{\lambda_X} \int_{-\infty}^{\lambda_X} \varphi_2 (s,t,\rho_X(h)) \int_{\frac{\lambda_Z - \sqrt{m}s}{\sqrt{1-m}}}^\infty \int_{\frac{\lambda_Z - \sqrt{m}t}{\sqrt{1-m}}}^\infty \varphi_2(\widetilde s, \widetilde t, \rho_Y(h)) \, \mathrm{d} \widetilde s \, \mathrm{d} \widetilde t \, \mathrm{d} s \, \mathrm{d} t
 \end{equation}
 and
 \begin{equation}\label{eq:C12}
     C_{12}(h) = \int_{-\infty}^{\lambda_X} \int_{\lambda_X}^{\infty} \varphi_2(s,t,\rho_X(h))\left( 1 - \Phi \left( \frac{\lambda_Z - \sqrt{m}t}{\sqrt{1-m}}\right)\right) \mathrm{d} t
 \end{equation}
 for each $h > 0$, relate the covariance functions $\rho_X$ and $\rho_Y$ of the random fields $X$ and $Y$ as well as the parameter $m$ to the two-point coverage probability functions $C_{22}$ and $C_{12}$. Then, we fix the parameter $m$, and compute an estimator $\widehat \lambda_Z$ for $\lambda_Z$ by Equation~\eqref{eq:epsilon2}. Here we plug in $\widehat \lambda_X$, which is obtained by Equation~\eqref{eq:epsilon1}, for $\lambda_X.$ In a further step, we plug in the estimators $\widehat \lambda_X, \widehat \lambda_Z$ and the non-parametric estimator $\widehat \rho_X$, computed by means of Equation~\eqref{eq:C11}, into Equation~\eqref{eq:C22}. Doing so, a non-parameteric estimator $\widehat \rho_Y$ for $\rho_Y$ is determined. Having estimated all the parameters for each fixed $m$ in the range of $0, 0.005, 0.010, 0.015, \ldots, 1,$ we plugin the obtained estimators into Equation~\eqref{eq:C12} and minimize the distance between the right-hand side of Equation~\eqref{eq:C12} and the estimator $\widehat C_{12}$ computed from image data. For numerical details of the described computation of estimators, we refer to Section~3 in~\cite{n.2019d}. Considering the final non-parameteric estimator $\widehat \rho_Y$, it is suitable to assume \begin{equation}\label{eq:choice_rhoY}
     \rho_{Y}(h) = \exp(- (\alpha_Y h)^2)
 \end{equation} for each $h > 0$, where $\alpha_Y > 0$ is a model parameter. As in the case of $\alpha_X$, an estimator $\widehat \alpha_Y$ for $\alpha_Y$ is computed by the method of least-squares. The values of the computed estimators are provided in Table~\ref{tab:estimated_parameters}.

 \begin{table}[h]
	\centering
	\caption{Model parameters estimated from tomographic 3D image data.}\label{tab:estimated_parameters}
	\begin{tabular}{ccccccc}
		$\widehat \lambda_X$ & $\widehat \lambda_Z$ & $\widehat \alpha_X$ & $\widehat \alpha_Y$ & $m$ \Tstrut \Bstrut \\ \hline 
		-0.0371 & -0.4886 & $16.6~\upmu\mathrm{m}^{-1}$ & $26.5~\upmu\mathrm{m}^{-1}$ & 0.15 \Tstrut
	\end{tabular}
\end{table}

\section{Results and discussion}\label{sec:resuls_and_discussion}

In this section, the main results of stochastic 3D modeling of nanostructured NVP/C-particles are discussed. First, we validate the proposed stochastic 3D struture model in Section~\ref{sec:validation_stochastic_model}. Second, in Section~\ref{sec:predictive_simulation}, the validated model is used for a simulation-based prediction on how the amount of carbon influences the morphology and thus effective properties of the nanostructured active material particles.

\subsection{Model validation}\label{sec:validation_stochastic_model}

For validating the stochastic structure model, we simulate three realizations of the calibrated model with the parameters shown in Table~\ref{tab:estimated_parameters} with a cubic voxel size of 16.6 nm$^3$. The nanostructures are realized within cubes of side length $4.98~\upmu \mathrm{m}$. Figure~\ref{fig:model_data_comparison} shows a comparison between a segmented HR-SEM image and a 2D cross-section of a model realization. Here one can observe a good visual fit, which is quantitatively underpinned by comparing morphological descriptors. 

\begin{figure}[h]
    \centering
    \begin{tabular}{ccc}
        \includegraphics[width = 0.32\textwidth]{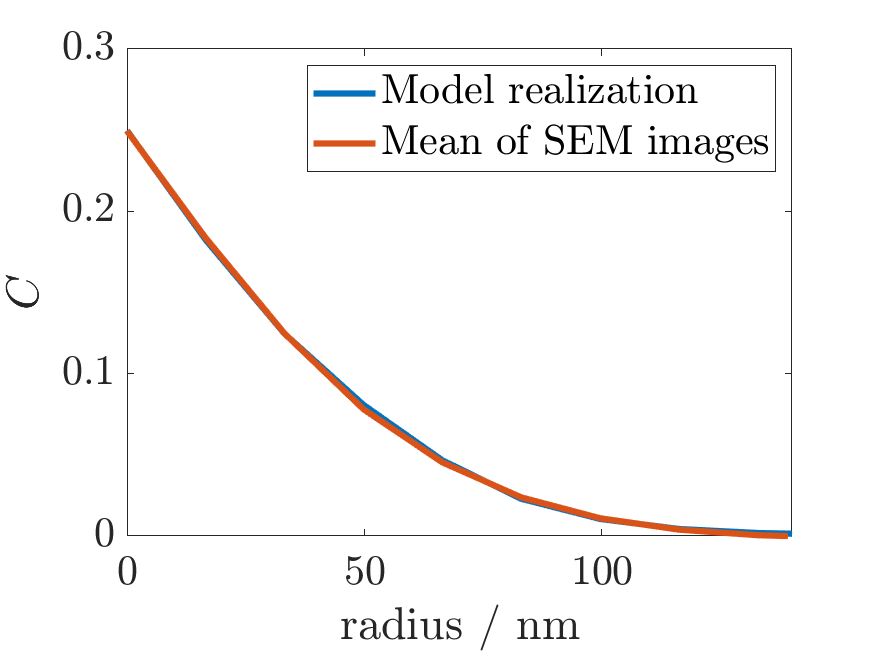} & 
        \includegraphics[width = 0.32\textwidth]{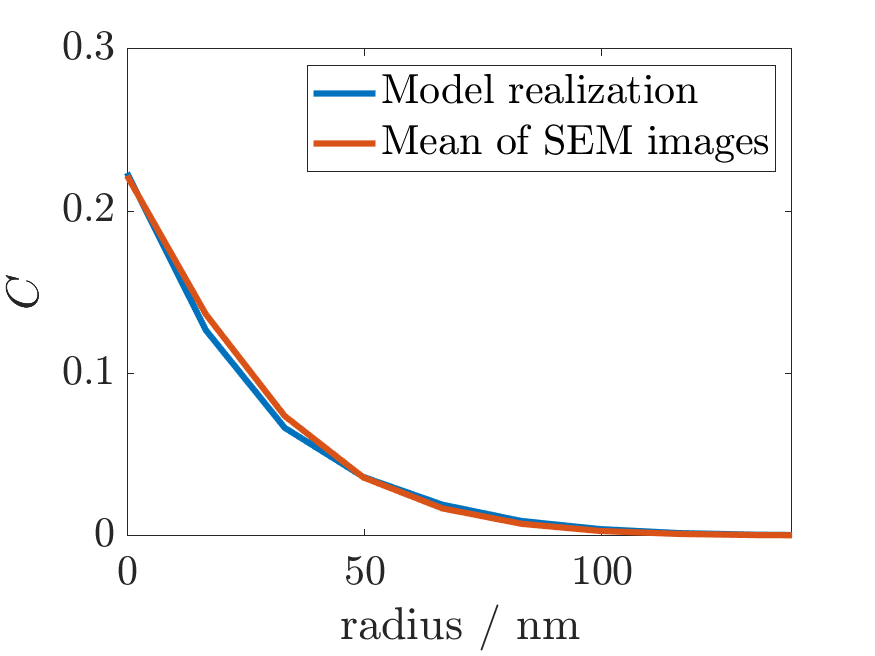} & 
        \includegraphics[width = 0.32\textwidth]{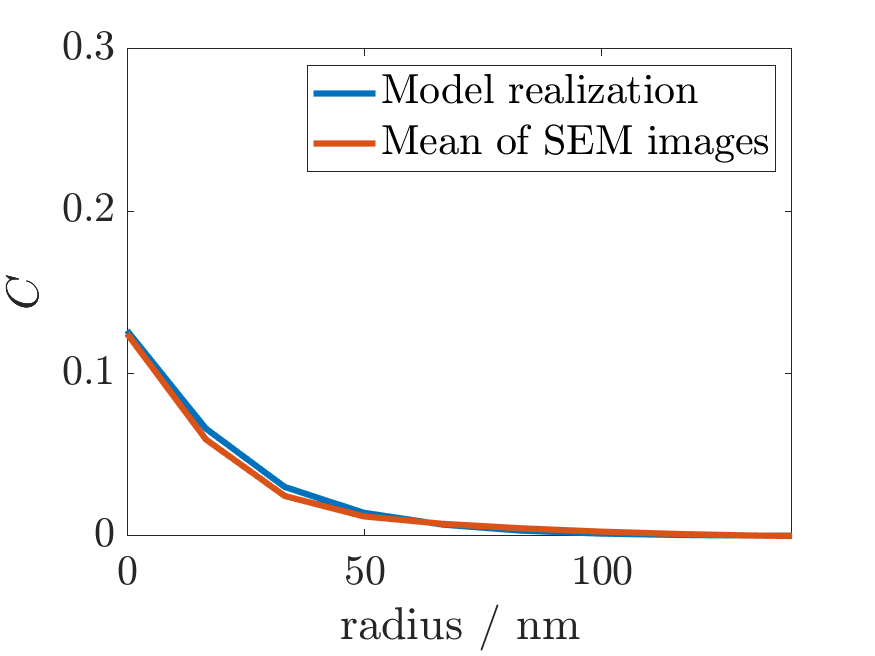}\\
        (a) & (b) & (c) \\
         \includegraphics[width = 0.32\textwidth]{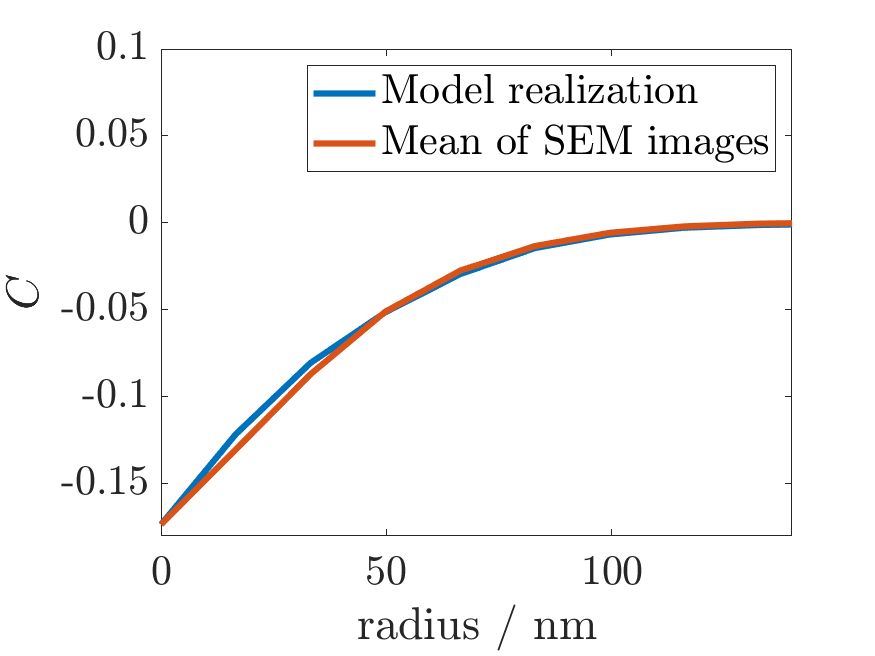} & 
        \includegraphics[width = 0.32\textwidth]{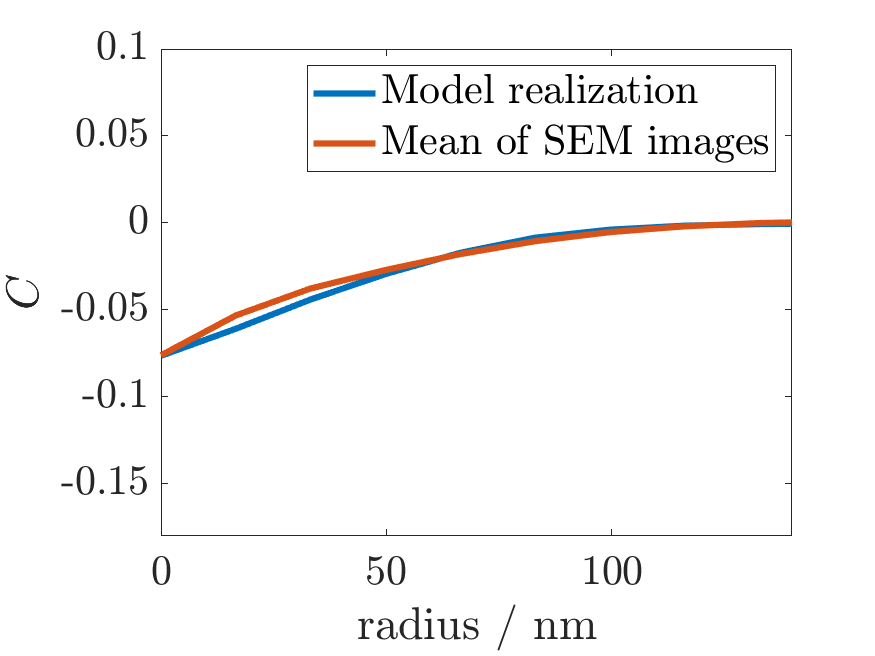} & 
        \includegraphics[width = 0.32\textwidth]{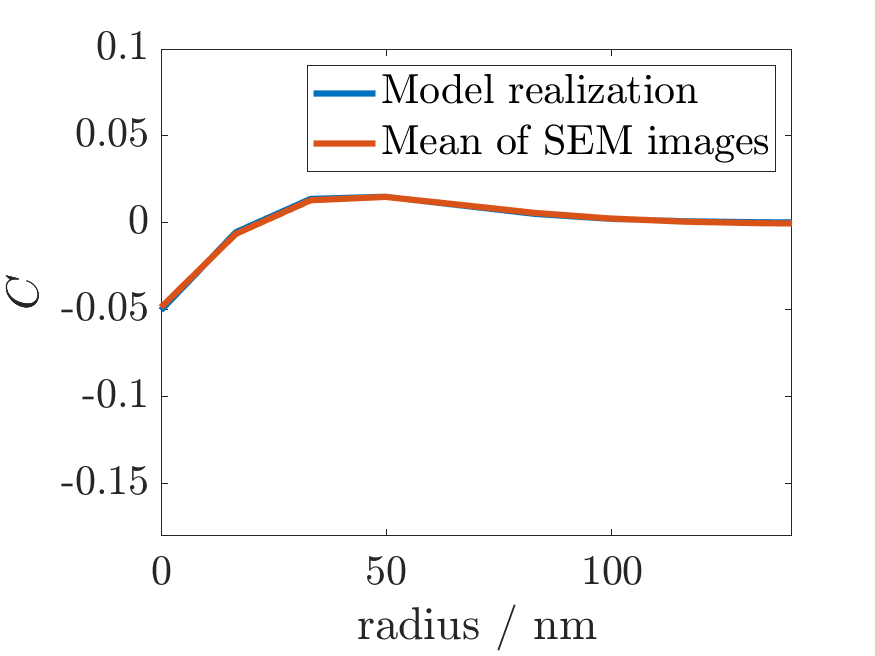} \\
        (d) & (e) & (f)
    \end{tabular}
    \caption{Centered two-point coverage probability functions $C_{ij} - \varepsilon_i \varepsilon_j$ estimated from HR-SEM images and simulated data. The cases $i=j=1$~(a), $i=j=2$~(b), $i=j=3$~(c), $i=1$ and $j=2$~(d), $i=1$ and $j=3$~(e) as well as $i=2$ and $j=3$~(f) are shown. Recall that $\Xi_1$, $\Xi_2$, $\Xi_3$ represent NVP, carbon, and pores, respectively. }
    \label{fig:Two-point_coverage}
\end{figure}

For the quantitative validation, we consider the two-point coverage probability functions introduced in Section~\ref{sec:model_fitting} as well as the 2D continuous phase size distributions~\cite{holzer.2013b}. The two-point coverage probability functions are shown in Figure~\ref{fig:Two-point_coverage}. The functions are centered here, i.e., we show the functions $C_{ij}(h) - \varepsilon_i \varepsilon_j$, such that all functions asymptotically tend to 0. It is evident that we have a nearly perfect fit of the model with respect to these descriptors. Even if these functions have been used for model calibration, the good accordance can be considered as model validation regarding the following two aspects. First, Figure~\ref{fig:Two-point_coverage} (a) shows the consistency between FIB-SEM images with a voxel size of $16~\mathrm{nm} \times 16~\mathrm{nm} \times 20~\mathrm{nm}$ and the SEM images with a pixel size of $5$~nm. Recall from Section~\ref{sec:model_fitting} that the correlation function $\rho_X$, which determines $C_{11}$ is estimated based on FIB-SEM data, while the Figure~\ref{fig:Two-point_coverage} (a) compares $C_{11}$ computed from model realizations with $C_{11}$ estimated from the three HR-SEM images which are used to quantify the spatial distribution of carbon and pores. Second, the results shown in Figure~\ref{fig:Two-point_coverage} justify the particular choice of the  covariance functions $\rho_X$ and $\rho_Y$ given in Equations~\eqref{eq:choice_rhoX} and~\eqref{eq:choice_rhoY}, which both depend only on one model parameter. It is not clear \emph{a priori} that this choice allows to fit all two-point coverage probability functions as well as it is shown in Figure~\ref{fig:Two-point_coverage}.

\begin{figure}
    \centering
    \begin{tabular}{ccc}
         \includegraphics[width = 0.32 \textwidth]{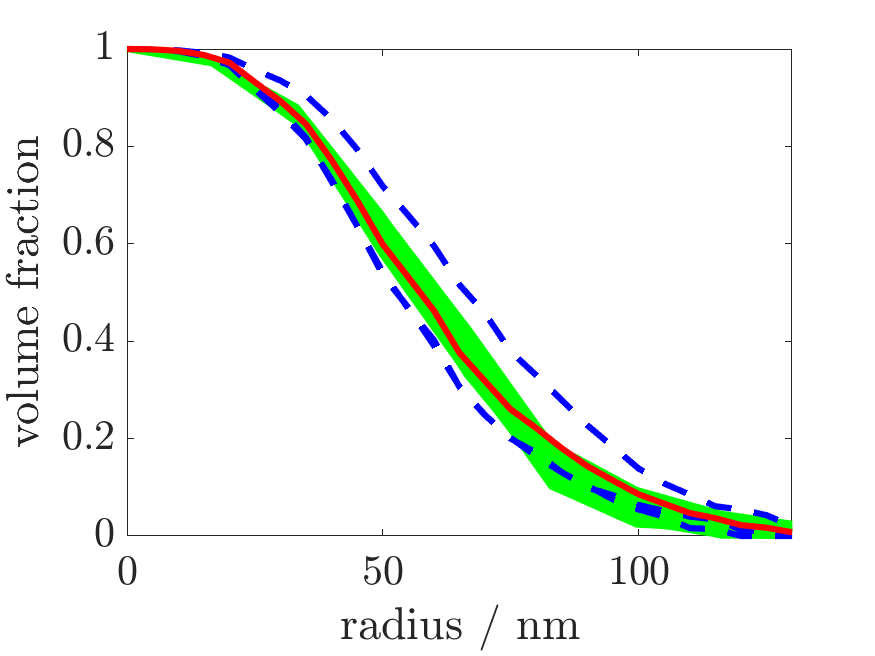} &  \includegraphics[width = 0.32 \textwidth]{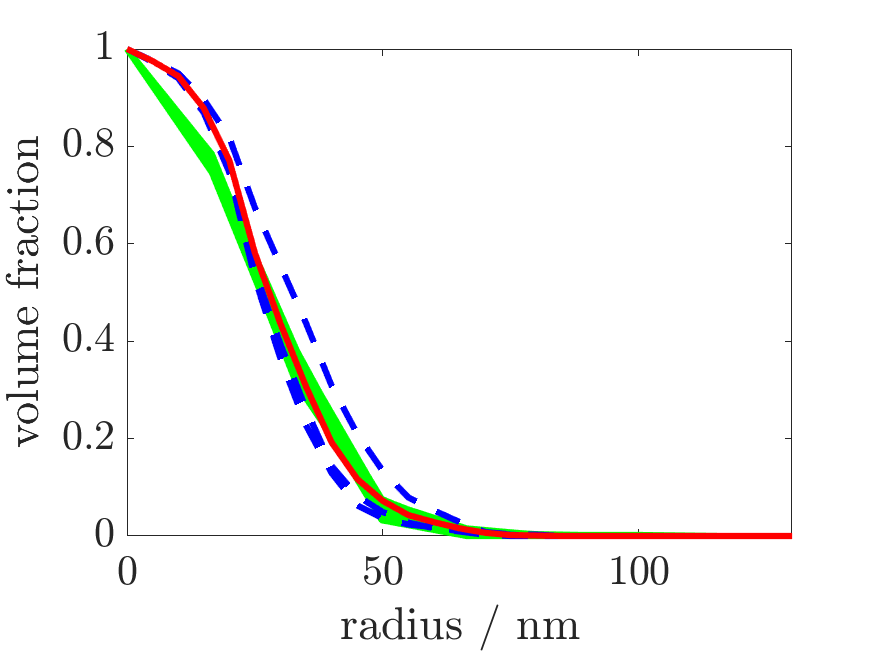} & 
         \includegraphics[width = 0.32 \textwidth]{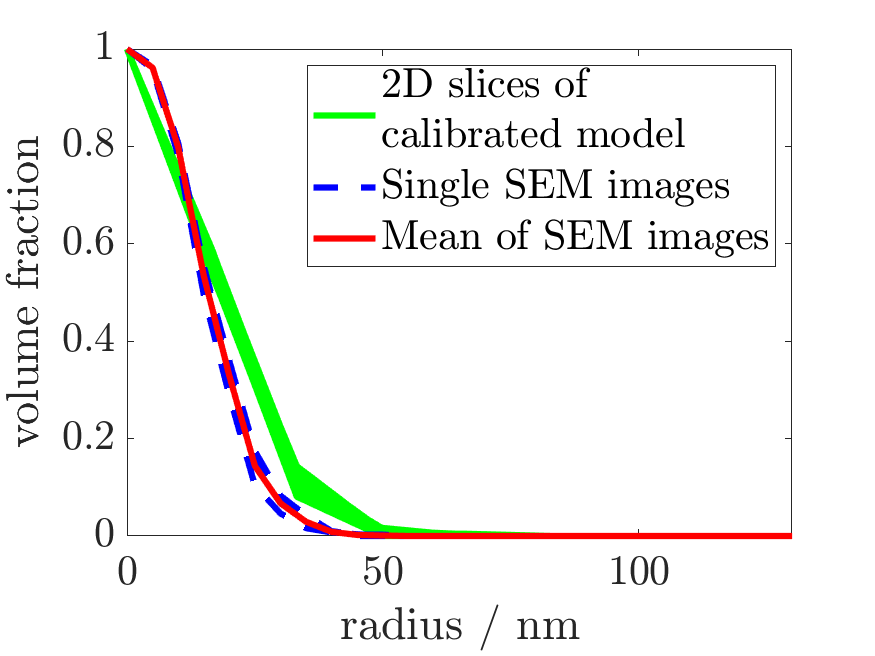}\\
         (a) & (b) & (c) 
    \end{tabular}
    \caption{Comparison of 2D continuous phase size distributions between HR-SEM images and simulated data. The green curves represent the continuous phase size distributions computed for each slice of the simulated 3D data. The blue dotted lines show the  continuous phase size distributions of the three HR-SEM images, the mean of which is represented in red. Results are shown for NVP (a), carbon (b) and pores (c).}
    \label{fig:2D_CPSD}
\end{figure}

Moreover, the 2D continuous phase size distribution of model realizations and HR-SEM images are compared in Figure~\ref{fig:2D_CPSD}. The 2D continuous phase size distribution $\Psi_i:[0,\infty) \rightarrow [0,1]$ of the phase $\Xi_i$, $1 \leq i \leq 3$ is defined by 
\begin{equation}\label{eq:CPSD}
    \Psi_i(h) = \frac{1}{\varepsilon_i \nu_2(W)} \E \left[ \nu_2\left(((\Xi \cap W) \ominus D(o,h)) \oplus D(o,h) \right) \right],
\end{equation}
where $W \subset \R^2$ with two-dimensional Lebesgue measure $\nu_2(W)$ and $D(o,h)$ denotes the disk centered at the origin $o \in \R^2$ with radius $h > 0.$ In other words, the continuous phase size distribution of $\Xi_i$ at $h > 0$ is the volume fraction of the morphological opening~\cite{soille.2003} of $\Xi_i$ with a disk of radius $h > 0$ as structuring element, which is normalized by the volume fraction of $\Xi_i$ itself. Note that this opening can be considered as the subset of $\Xi_i$ which can be covered by a union of disks $D(o,h),$ where each disk is completely contained in $\Xi_i$. The continuous phase size distribution carries the same information as the granulometry function in mathematical morphology~\cite{matheron.1975, jeulin.2021}, which is a widely used descriptor measuring the size distribution of complex structures. We estimate the continuous phase distribution at $h > 0$ from image data by estimating the volume fraction of the morphological opening of $\Xi_i$ with structuring element $D(o,h).$ Edge effects are avoided by means of minus-sampling as described in Section~4.7.2 of~\cite{chiu.2013}. Figure~\ref{fig:2D_CPSD} shows that the mean 2D continuous phase distribution of the HR-SEM images shows a rather similar behavior than the 2D continuous phase size distributions of the cross-sections of the simulated data. In particular for NVP, the fit is good in the sense that the mean curve computed from the HR-SEM images is completely contained within the domain in which the curves computed from the cross-sections of FIB-SEM data are located. Even if slight deviations are observed for carbon and pores, the model captures the continuous phase size distribution well. Note that these results validate the choice of the stochastic structure model described in Section~\ref{sec:model_description} since continuous phase size distributions have not been used for model calibration. Note that a good fit of the two-point coverage distributions does not necessarily imply a good fit with respect to the continuous phase size distributions, see the discussion in Section~4.3 of~\cite{n.2019d}.

\subsection{Predictive simulation: Varying the volume fraction of carbon}\label{sec:predictive_simulation}

Having validated the stochastic model, it can be used for predictive simulation of virtual, but realistic NVP/C electrodes where the volume fraction of carbon is varied for a constant NVP volume fraction of 0.5148, which has been estimated from image data (Section~\ref{sec:model_fitting}). Based on these virtual nanostructures, effective transport properties, i.e., effective conductivity and effective diffusivity of the three phases respectively, are predicted for different volume fractions of carbon by means of the empirical relationships derived in~\cite{stenzel.2016, n.2020}. Recall from Section~\ref{sec:materials_imaging} that the volume fraction of carbon means the volume fraction of carbon and nanopores within carbon that are not resolved in the segmentation of HR-SEM data. Under the assumption that computations in Section~\ref{sec:synthesis} correctly reflect the mere volume fraction of carbon without nanopores, we propose the following recomputation. The mere volume fraction of carbon in the complement of NVP computed in Section~\ref{sec:synthesis} is 0.28, while--due to possible nanopores that are not resolved in HR-SEM--the same fraction obtained from HR-SEM data is 0.69 (Section~\ref{sec:model_fitting}). The volume fraction of carbon without nanopores can then be computed by multiplying the volume fractions of carbon with small nanopores by $0.28/0.69 = 0.41.$

\begin{figure}[h]
    \centering
    \begin{tabular}{ccc}
     \includegraphics[width = 0.33\textwidth]{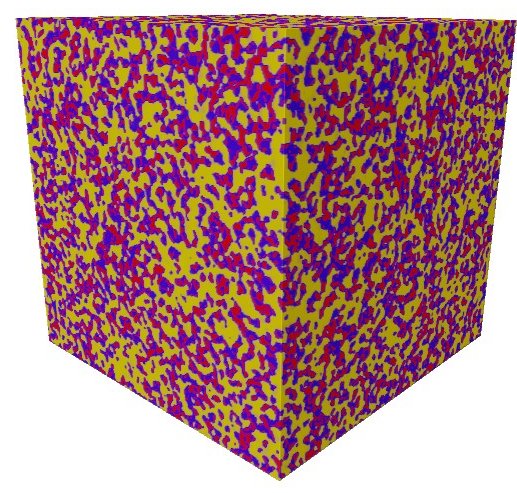} &
    \includegraphics[width = 0.33\textwidth]{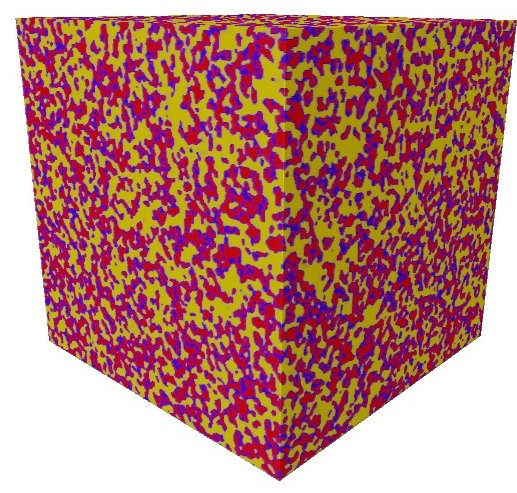}&
     \includegraphics[width = 0.33\textwidth]{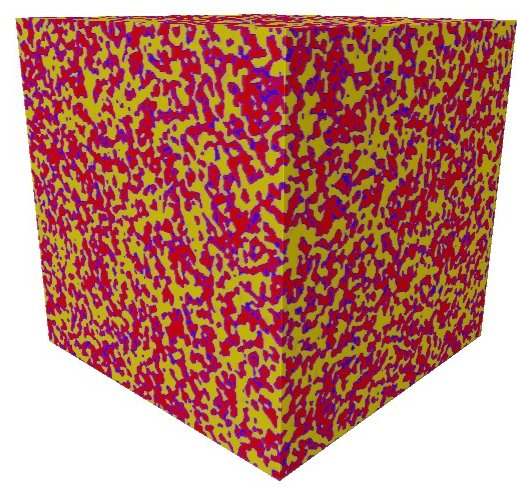}\\
     (a) & (b) & (c) 
\end{tabular}
    \caption{3D rendering of virtual NVP/C nanostructures with 20~\% (a), 30~\% (b) and 40~\% (c) volume fraction of carbon (including small nanopores). The size of each structure is $4.98~\upmu \mathrm m \times 4.98~\upmu \mathrm m \times 4.98~\upmu \mathrm m.$ NVP is represented in yellow, carbon in red and pores in blue.}
    \label{fig:3Dvis_predictive_sim}
\end{figure}

For this purpose, we keep all model parameters fix except of $\lambda_Z$, which controls the volume fraction of carbon denoted by $\varepsilon_2$. We consider $\varepsilon_2 \in \{ 0.150, 0.175, \ldots, 0.400 \}$. For a fixed value of $\varepsilon_2$, we plug-in $\varepsilon_2$, $\lambda_X$, $m$ into Equation~\eqref{eq:epsilon2}. Due to the monotonicity of the right-hand side in $\lambda_Z,$ we can numerically solve for $\lambda_Z$ using the method of bisection. This means, that for each predefined value of the volume fraction of carbon $\varepsilon_2,$ we obtain the full set of model parameters, which can then be used to generate virtual NVP/C nanostructures. We generate three realizations for each $\varepsilon_2 \in \{ 0.150, 0.175, \ldots, 0.400 \}$. Corresponding 3D visualizations are represented in Figure~\ref{fig:3Dvis_predictive_sim}. It is important to note that these predictive simulations are based on the simplifying assumption that varying the amount of carbon does only influence the model parameter $\lambda_Z.$ A validation of this assumption would require further image data of NVP/C particles synthesized with different amounts of carbon.

For the virtual structures, morphological descriptors, which are electrochemically important, such as the pairwise interface area between the phases, mean geodesic tortuosity quantifying the length of shortest transportation paths as well as constrictivity which is a measure for botteleneck effects are computed. In combination with the volume fractions, the latter two descriptors are used to predict the effective transport properties of the virtual structures. Recall that the volume fractions of the three phases are adjusted by the parameters of the stochastic model. 

\begin{figure}
    \centering
    \begin{tabular}{ccc}
     \includegraphics[width = 0.32\textwidth]{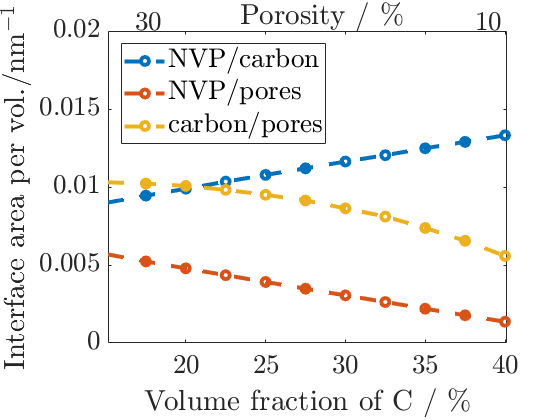} &
    \includegraphics[width = 0.32\textwidth]{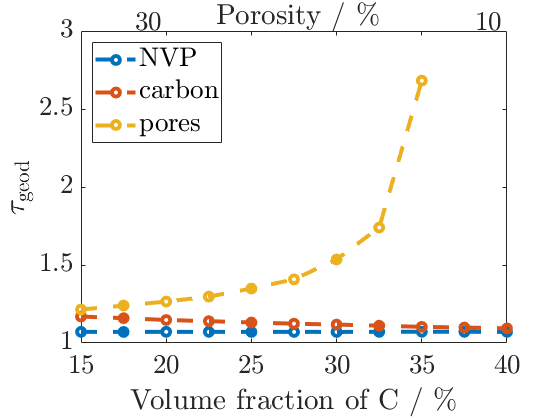}&
     \includegraphics[width = 0.32\textwidth]{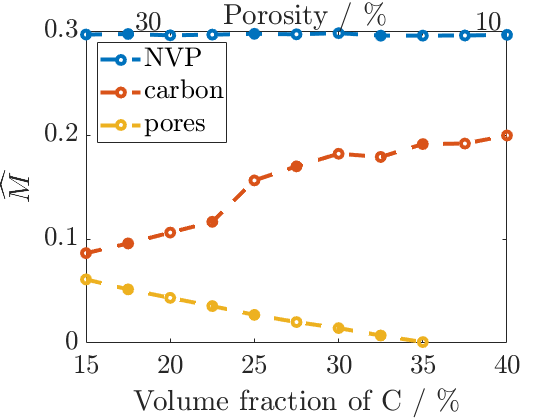}\\
     (a) & (b) & (c) 
\end{tabular}
    \caption{Influence of the volume fraction of carbon (including small nanopores) on the pairwise interface area per unit volume~(a), mean geodesic tortuosity~(b), and the $M$-factor predicted by Equation~\eqref{eq:prediction_formula_Mfactor}~(c). The results are shown for a fixed NVP volume fraction of 0.5148.}
    \label{fig:predictive_sim_quantitative}
\end{figure}

For estimating the interface areas from image data, we exploit the algorithm presented in~\cite{schladitz.2007} to compute the surface area of each phase. From this information, we can compute the pairwise interface areas using Equation (3) of~\cite{n.2018e}. Figure~\ref{fig:predictive_sim_quantitative}a shows the pairwise interface areas per volume dependent on the volume fraction of carbon. For the interface area per volume between NVP and carbon and between NVP and pores we can observe a linear increase and decrease, respectively, for an increasing volume fraction of carbon. The relationship between the interface area of carbon and pores and the volume fraction of carbon, on the contrary, is not linear. Keeping in mind that the volume fraction of NVP is fixed in our simulations, the behavior shown in Figure~\ref{fig:predictive_sim_quantitative}a is reasonable. The higher the carbon content is in the nanostructured particles, the less pore space is there and thus the interface area per volume of carbon and pores will have a maximum for volume fraction of carbon between $0 \%$ and $20 \%$.

Mean geodesic tortuosity denoted by $\tau_\mathrm{geod}$ of a phase is defined 
as the quotient of the expected length of shortest paths through the considered phase,  divided by the thickness of the material~\cite{n.2020}. Note that mean geodesic tortuosity is a purely morphological descriptor. For an overview about different notions of tortuosity, we refer to~\cite{clennell.1997, ghanbarian.2013, holzer.2023}. A formal definition of mean geodesic tortuosity in the framework of random closed sets can be found in~\cite{n.2019b}. Figure~\ref{fig:predictive_sim_quantitative}b shows how mean geodesic tortuosity of the three phases depends on the volume fraction of carbon. As expected, a slight decrease of mean geodesic tortuosity is observed for an increasing volume fraction of carbon. The values for NVP and carbon are low. The mean geodesic tortuosity of the pore space, which gives information about the length of shortest pathways for ions being transported through the pores, is more interesting. While $\tau_\mathrm{geod}$ of the pore space is similar to the values of NVP and carbon at a carbon volume fraction of $15~\%$, the values of the pore space are strongly increasing with an increasing volume fraction meaning that the pore space is reduced. For volume fractions of carbon between $35~\%$ and $37.5~\%,$ there is the percolation threshold of the pore space. At these or higher volume fractions of carbon, there are no pathways through the pore space from one side to the other in the cube of side length $4.98~\upmu m$.

An important quantity determining the effective properties of aggregated active material particles are the effective transport properties, namely effective diffusivity and effective conductivity, of the pore space, NVP and carbon, respectively. The $M$-factor, see Section~5.2 in~\cite{holzer.2023}, is defined as the ratio of effective and intrinsic diffusivity (and analogously conductivity). Together with constrictivity $\beta$, the information contained in the volume fraction and mean geodesic tortuosity can be used to predict the $M$-factor. Constrictivity $\beta$ is a descriptor for the strength of bottlenecks effects, given as squared ratio of the width of the typical bottleneck obtained from  simulated mercury intrusion porosimetry over the median of the 3D continuous phase size distribution, see~\cite{holzer.2013b, n.2019b} for details. An empirically derived prediction $\widehat M$ for the $M$-factor~\cite{stenzel.2016} is given by 
\begin{equation}\label{eq:prediction_formula_Mfactor}
    \widehat M = \frac{\varepsilon^{1.15} \, \beta^{0.35}}{\tau_{\mathrm{geod}}^{4.39}}.
\end{equation}
Note that this prediction formula has been particularly validated for nanostructured NMC particles by finite element modeling in~\cite{neumann.2023}. The predictions of $M$-factors are shown in Figure~\ref{fig:predictive_sim_quantitative}c. Since the model parameters controlling NVP are not influenced in the simulation study, the corresponding predicted $M$-factor does not change. For the pore space, we observe a linear decrease of the $M$-factor until the percolation threshold is reached. The $M$-factor of carbon is, as expected, monotonously increasing in the range between $0.1$ and $0.2$. This means that increasing the volume fraction of carbon from $15~\%$ to $40~\%$ allows for doubling the value of the effective conductivity. 

\section{Conclusion}\label{sec:conclusion}

In the present paper, we have used stochastic 3D modeling to generate virtual structures which are statistically similar to the nanostructure of NVP/C particles used as active material for sodium-ion batteries. Model calibration has been performed based on image data. For this purpose, experimental data sets have been obtained with FIB-SEM, high-resolution SEM, EDX and TEM. While the morphology of NVP can be appropriately reconstructed from FIB-SEM data, high-resolution SEM is required to resolve the morphology of carbon and filled pores. The Pluri-Gaussian model introduced in~\cite{n.2019d} is calibrated using both, FIB-SEM and HR-SEM data. Here it is important to note that statistical techniques are used which allow to estimate these parameters of the 3D model, which concern the spatial distribution of carbon and pores, merely based on 2D SEM data. The good agreement between model and data in terms of morphological descriptors which are not used for model calibration validates the choice of the Pluri-Gaussian model. Finally, the parameters of the stochastic model are systematically varied for virtual scenario analysis. In particular, the influence of the volume fraction of carbon on electrochemically important descriptors such as the area of interfaces per unit volume and effective transport properties is quantitatively investigated. These values are important input parameters for modeling and simulation of effective properties of nanostructured NVP/C active material particles within the composite electrode, which strongly influence the performance of the battery.

\section*{Acknowledgements}
 
The Focused Ion Beam Center UUlm is acknowledged for enabling FIB/SEM and EDX analysis. The Central Facility for Electron Microscopy (Clarissa Read) is thanked for TEM imaging. This work contributes to the research
performed at CELEST (Center for Electrochemical Energy Storage Ulm -
Karlsruhe) and was funded by the Deutsche Forschungsgemeinschaft
(DFG, German Research Foundation) under Germany’s Excellence
Strategy – EXC 2154 – Project number 390874152 (POLiS Cluster of
Excellence).

\newpage

\section*{Supplementary information}
\setcounter{page}{1}
\setcounter{figure}{0}
 \renewcommand\thefigure{S\arabic{figure}}

\begin{figure}[h]
    \centering
    \includegraphics[width = \textwidth]{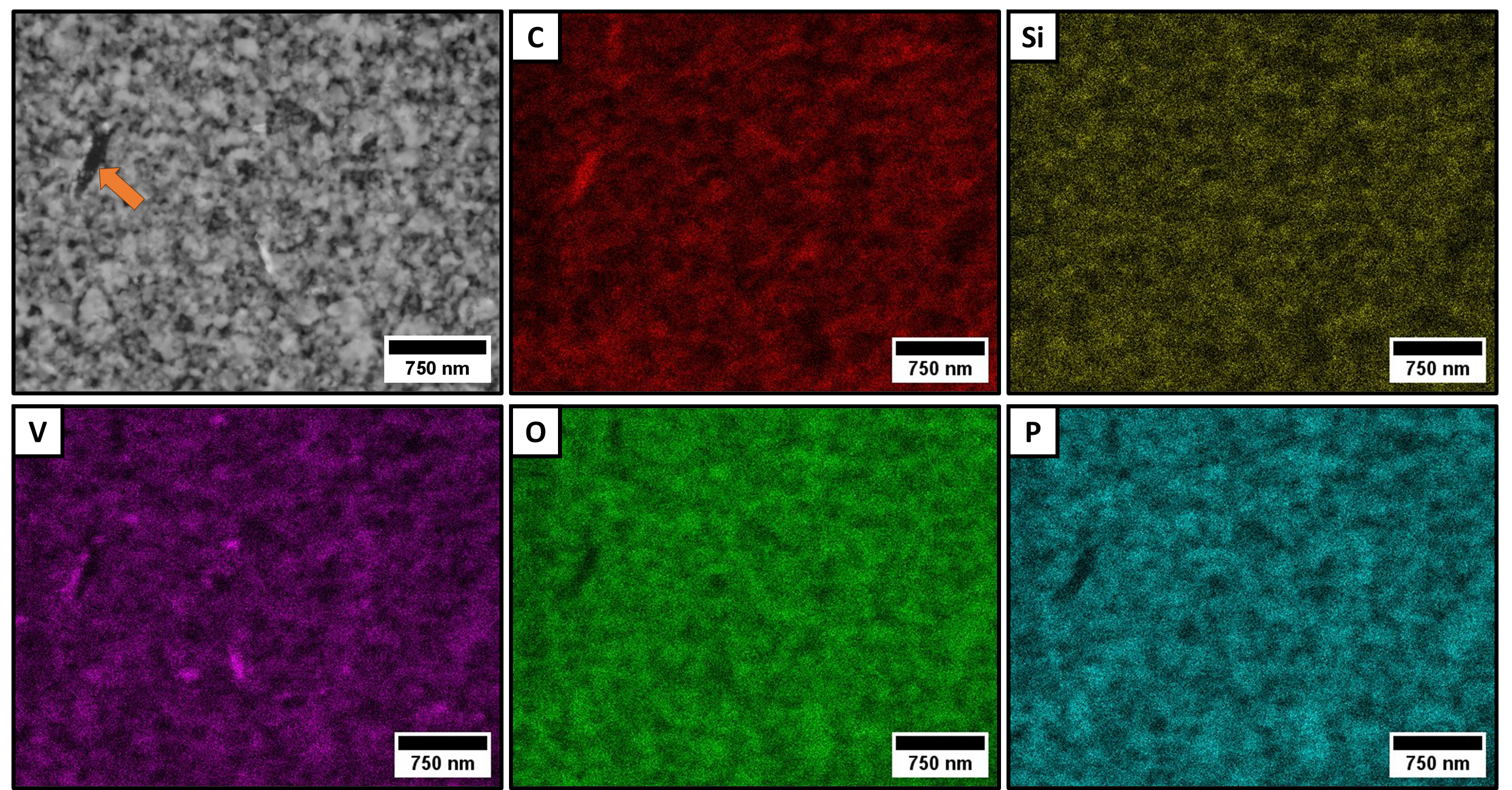}
    \caption{SEM image and element maps of C, Si, V, O and P of the prepared lamella to assist the assignment of the gray values to the different phases present within the material. The prominent carbon particle (orange arrow) was further investigated by TEM, see Figure~\ref{fig:tem}.}
    \label{fig:EDX}
\end{figure}

\begin{figure}[h]
\begin{tabular}{cc}
     \includegraphics[width = 0.33\textwidth]{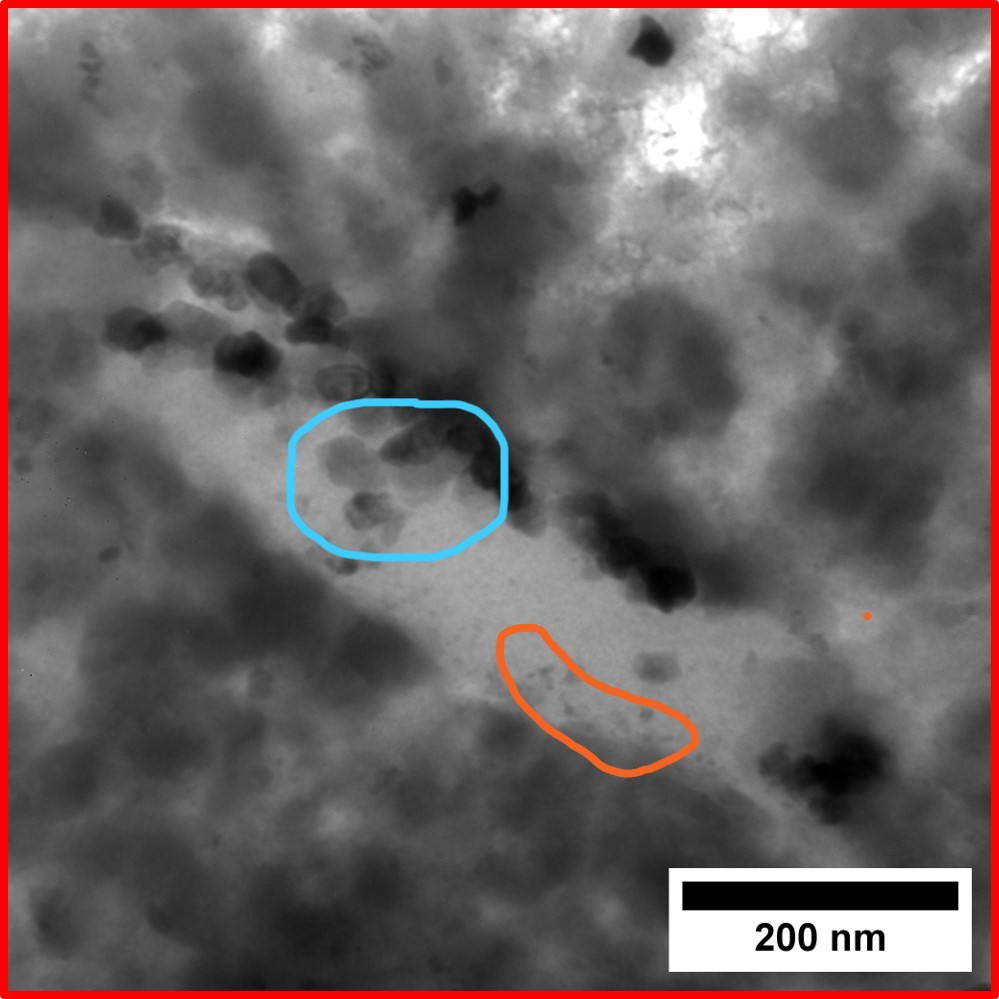} & \includegraphics[width = 0.375 \textwidth]{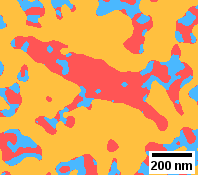} \\
     (a) & (b)
\end{tabular}
\caption{TEM image of a prepared lamella from the NVP secondary particle zoomed to the carbon particle within the NVP/C particle visible in the EDX analysis shown in Figure \ref{fig:EDX}. Dark areas correspond to the NVP phase, gray areas to the silicone resin and bright areas to the carbon phase (a). For comparison, the corresponding part of the segmented HR-SEM image shown in Figure~\ref{fig:EDX} is visualized. The blue and orange regions indicate that the amount of pores within carbon are underestimated in the segmentation of HR-SEM data (b).}\label{fig:tem}
\end{figure}

As supplementary information, results of EDX (Figure~\ref{fig:EDX}) and TEM (Figure~\ref{fig:tem}) measurements are visualized. The detailed analysis of a carbon particle by TEM shows two effects leading to the overestimation of the volume fraction of carbon while underestimating the porosity when using HR-SEM images as a basis.  
 First, the amount of nanopores is underestimated in the region encircled in blue. While TEM data suggests an interconnected pore spaces there, three separated pore regions are segmented from SEM data. Moreover, there are rather small nanopores visible in the region encircled in orange. These small pore regions are not completely represented in the segmentation of HR-SEM data. 

\end{document}